\documentclass[oneside,a4paper,reqno]{amsart}
\usepackage{amsmath}
\usepackage{amssymb}
\usepackage{xspace}
\usepackage[mathscr]{eucal}
\usepackage{graphics}

\newcommand{\ket}[1]{\ensuremath{		\left| #1 \right> 
																																			  }
																									}

\newcommand{\coh}[3]{\ensuremath{		\left( #1, #2 \right)_{#3} 
																																			  }
																									}

\newcommand{\overlap}[2]{\ensuremath{ 
																								\left \langle    #1 \vphantom{#2 } \,
                        \right| \left.   #2 \vphantom{#1}
                        \right \rangle
                        									}
                     }
\newcommand{\boverlap}[2]{\ensuremath{ 
																								\bigl \langle #1 \, \bigr| \bigl. #2 \bigr \rangle
                        									}
                     }

\newcommand{\RErr}{\ensuremath{     \Delta_{\mathrm{ei}}  }}

\newcommand{\vb}{v_{\mathrm{B}}}
\newcommand{\bvb}{\bar{v}_{\mathrm{T}}}
\newcommand{\bvl}{\bar{v}_{\lambda}}

\DeclareMathOperator{\imaginary}{Im}

\DeclareMathOperator{\sech}{sech}
\begin{document}
\begin{titlepage}
\begin{center}
\bfseries
GENERIC BOHMIAN TRAJECTORIES OF AN ISOLATED PARTICLE
\end{center}
\vspace{1 cm}
\begin{center}
D M APPLEBY
\end{center}
\begin{center}
Department of Physics, Queen Mary and
		Westfield College,  Mile End Rd, London E1 4NS, UK
 \end{center}
\vspace{0.5 cm}
\begin{center}
  (E-mail:  D.M.Appleby@qmw.ac.uk)
\end{center}
\vspace{0.75 cm}
\begin{center}
  QMW--PH--99--06
\end{center}
\vspace{1.25 cm}
\begin{center}
\textbf{Abstract}\\
\vspace{0.35 cm}
\parbox{10.5 cm }{ 
     The generic Bohmian
     trajectories are calculated for an isolated particle in 
     an approximate energy eigenstate, for an arbitrary one-dimensional
     potential well.
     It is shown, that the necessary and sufficient condition
     for there to be a negligible probability of the trajectory 
     deviating significantly from the classical trajectory
     at any stage in the motion
     is, that the state be a  narrowly localised wave
     packet.    The properties of the 
    Bohmian trajectories are  discussed
    in relationship to the theory of 
     retrodictively optimal 
    simultaneous measurements of position and momentum
    which was presented in several previous papers.
    It is shown that
    the Bohmian velocity at
     $x$ is
    the expectation value of the velocity which 
    would be observed at $x$, if one
    were to make a retrodictively optimal simultaneous 
    measurement of $x$ and $p$, in the limit as the error in the 
    measurement of $x$ tends to zero.  This explains 
    the tendency of the Bohmian particle to behave in a highly
    non-classical manner.  It also
    explains why the trajectories
    in the interpretation 
    recently proposed by Garc\'{i}a de Polavieja tend to be
    much more nearly classical in the limit of large quantum 
    number.
    The implications for other trajectory interpretations are 
    considered.
                 }
\end{center}
%\vspace{1 cm}
%\begin{center}
%Report no.  QMW-PH-98-12
%\end{center}
\end{titlepage}
%\title{Generic Bohmian Trajectories of an
%Isolated Particle}
%\author{D.M.~Appleby}
%\date{\today}
%\maketitle
\section{Introduction}  
\label{sec:  Introduction}
This is the first of two papers in which we investigate the classical
limit in the Bohm interpretation of quantum
mechanics~\cite{Bohm1,Bohm2,Holland1}.  It is well known that the Bohmian
trajectories can be highly non-classical.  We are interested in the
question, whether this is true of the macroscopic bodies of our ordinary
experience.  

Different views have been expressed in the literature.  
Bohm and Hiley~\cite{Bohm2}
have argued that the Bohm interpretation does successfully account for the
existence of an approximately classical level of phenomena due to the
effect of the electromagnetic radiation and other particles incident on a
macroscopic object such as a planet.  On the other hand,
Holland~\cite{Holland2} has argued that Bohm's theory may not be rich
enough to embrace the full variety of possible classical motions and, in
consequence, that it may not be a universal physical theory. 
Holland~\cite{Holland3} has gone on to propose an alternative trajectory
interpretation, which he hopes may prove more satisfactory in this
respect.

The issues raised by these authors are of some importance. There has been
much
discussion~\cite{Englert,Durr,Englert2,Dewdney,Aharanov,Scully,Griffiths}
of the fact that the Bohmian trajectory of a micro-object can, under
certain circumstances, be ``surreal''.   Although this behaviour is
highly counter-intuitive, it does not provide the grounds for a clear
logical objection since there is no actual conflict with experiment.  On
the other hand, it would create very serious difficulties for the
interpretation if it could be shown that, under the conditions of our
ordinary experience, the Bohmian trajectory of a macro-object can be
significantly different from the trajectory predicted by classical
mechanics.  This is because the Bohm interpretation is usually based on
what Fine~\cite{Fine} describes as an assumption of ``accessibility''.
That is, it is assumed that, at least in the case of a macro-object, the
position which is (as one would normally say) directly perceived closely
corresponds to the position which actually exists (\emph{modulo}
exceptional instances of hallucination \emph{etc.}).
Bell~\cite{Bell} makes the point with his usual vigour and clarity when he
says that, in the Bohm interpretation, the positions of macroscopic
objects, under the conditions under which we normally experience them, are very
far from being  ``hidden'':
\begin{quote}
Absurdly, such theories are known as `hidden variable' theories.  
Absurdly,
for there it is not in the wavefunction that one finds an image of the
visible world, and the results of experiments, but in the complementary
`hidden'(!) variables.
\end{quote}
It should be noted that it is not simply 
an instantaneous image that is needed. One also needs to
be able to assume that our memory traces, of the way in which a 
macroscopic
body appears to have moved in the past, closely correspond to the way in
which it actually moved.  In other words, one needs the whole trajectory
to be ``accessible'', and not just the instantaneous position.

It may be asked whether this assumption is strictly necessary.  If one
drops the assumption, then one is committed to the view that the actual
trajectory of a macroscopic body can be markedly and
systematically different from its apparent trajectory.  It might,
perhaps, be possible to reconstruct the Bohm interpretation along such
lines.  Indeed,  Page~\cite{Page} has made some definite proposals in this
connection.  However, as Page points out, one would then be making the
interpretation depend on profoundly difficult, and hitherto unresolved
questions regarding the nature of human consciousness.  Moreover, it is
hard to see what would be achieved by postulating the existence of
``beables'' of such a radically elusive kind.  An ontological
interpretation such as this has no obvious advantage over the Copenhagen
interpretation.  In short, although the assumption might, perhaps, not be
strictly necessary, dropping the assumption would involve the
interpretation in very considerable difficulties.  This is why the
questions discussed by Bohm and Hiley~\cite{Bohm2}, and by
Holland~\cite{Holland2}, are important.  If it should transpire
that the Bohmian trajectories of macroscopic objects are not typically
quasi-classical under the conditions of our ordinary experience, then the
assumption of accessibility would clearly not be justified.

In the sequel to this paper we will give some additional arguments
in support of  Bohm and Hiley's conclusion, that environmental effects
cause the Bohmian trajectory of a macroscopic object typically to become
quasi-classical.  However, before considering the effect of the
environment, it is natural  to ask what is the typical behaviour when the
body is isolated.  This is the question addressed in the present paper.

The paper is in two main parts.
The purpose of the first part (Sections~\ref{sec:  WKB} 
and~\ref{sec:  TimeAve}) is to 
investigate the  sense in which it is true, that the Bohmian trajectories
of an isolated body are \emph{generically} non-classical.

Consider a particle moving in an arbitrary one-dimensional potential well. 
If it is in an  energy eigenstate, then the Bohmian velocity is
zero.  However, it could be argued~\cite{Squires} that the significance
of this fact is somewhat unclear, since such states are
not typical.  One might argue that a macroscopic body is very unlikely to
be in an \emph{exact} energy eigenstate.  We are therefore led to consider
the case when the system is in an approximate energy  eigenstate, of the
form
\begin{equation}
  \ket{\psi} = \sum_{r=-\frac{\Delta n}{2}}^{\frac{\Delta n}{2}}
                  c_{r} \ket{\bar{n}+r}
\label{eq:  ApproxEState}
\end{equation}
where $\ket{n}$ denotes the $n^{\mathrm{th}}$ energy  eigenstate, with
energy $E_{n}$. Since we are interested in the classical limit we assume
that the state is highly excited, $\bar{n}\gg1$.  The fact that 
$\ket{\psi}$ is an approximate energy eigenstate means that
$\Delta n
\ll
\bar{n}$.   

We wish to establish the conditions which the coefficients
$c_{r}$ must satisfy in order to ensure that there is a negligible
probability of non-classical behaviour.
We begin, in Section~\ref{sec:  WKB}, by showing that the necessary and
sufficient condition for there to be a negligible probability of the
instantaneous Bohmian velocity being significantly different from the
classical velocity at any time during the motion is that
$\ket{\psi}$ is a  narrowly localised wave-packet.

This result does not entirely settle the question since the instantaneous
Bohmian velocity typically undergoes rapid fluctuations. 
Squires~\cite{Squires} has argued that, for the purposes of a comparison
with classical physics, the relevant quantity to consider
is, not the instantaneous velocity, but a suitable
time-average.  In Section~\ref{sec:  TimeAve} we calculate the time-averaged
velocity.  We show that the necessary and sufficient condition for there to be
probability~$\approx 1$ of the time-averaged velocity always being close
to the classical value is again, that  $\ket{\psi}$ is a  narrowly
localised wave-packet.

These results show that, for an isolated particle in
an approximate energy eigenstate, the Bohm interpretation produces
quasi-classical  trajectories in just those cases where no such
interpretation is needed (the conceptual difficulties which originally
led Bohm to propose his interpretation arise from the possible occurrence
of superpositions of macroscopically distinguishable states, as in the
paradigmatic instance of Schr\"{o}dinger's cat~\cite{Schrod}).  They
consequently show that the interaction with the environment plays an
essential role in the Bohm interpretation, just as it does in other
approaches to the problem of
interpretation~\cite{GellMann,Omnes,Joos,Zurek1,Zeh}.

In the second part of the paper (Sections~\ref{sec:  GdeP}
and~\ref{sec:  Others}) we investigate the underlying reasons for the
behaviour identified in the preceding Sections.  The Bohm interpretation,
although it was historically the first, and although it seems to be
mathematically the most straightforward, is by no means the only
interpretation in which the particles follow well-defined
trajectories~\cite{Holland3,Epstein,Bohm3,Bohm4,Nel1,Nel2,
Goldstein2,Vink,Singh1,Singh2,Singh3,Floyd1,Floyd2,Faraggi,
Polavieja,Bub,Suth,Ghir,Bac2,Potvin}. 
The interpretation proposed by Garc\'{i}a de Polavieja~\cite{Polavieja} is
particularly noteworthy from our point of view, since
it would appear to   produce the correct classical limit even in the case
of an isolated system, without any need to take into account the effect of
the environment.  This suggests that the counter-intuitive behaviour of
the Bohmian trajectories is not a general characteristic of every
possible trajectory interpretation, but is due instead to the particular
manner in which the Bohm interpretation has been constructed.  In 
Section~\ref{sec:  GdeP} we compare the two interpretations.  We argue
that the reason the trajectories in Garc\'{i}a de Polavieja's
interpretation tend to become quasi-classical in the limit of large
quantum number is connected with the fact that the phase space
distribution for this interpretation is the Husimi
function~\cite{Hus,Hil,Lee}, which  plays an important role in the
theory of simultaneous measurement
processes~\cite{Arthurs,Ali,Busch,Leon,Self4,Self3}.
We go on to argue that this also gives some insight into the reason why
the trajectories in the Bohm interpretation tend to be highly
non-classical (if the system is isolated).  In Section~\ref{sec: 
Others} we briefly consider the implications for other trajectory
interpretations.

\section{The Instantaneous Velocity}
\label{sec:  WKB} 
We consider a particle moving in one space dimension
under the influence of a potential
$V(x)$.  For simplicity we will assume that $V(x)$ has a single minimum,
and that the spectrum of the
Hamiltonian is purely discrete.  It would not be difficult to extend the
discussion to the case of more general potentials.  

Since we are interested in the limit of large quantum number it is
appropriate to use the WKB approximation. Let $\ket{n}$ be the
$n^{\mathrm{th}}$ energy eigenstate, and let $E_{n}$ be the corresponding
eigenvalue.  Let 
\begin{equation*}
  p_{n}(x) = \sqrt{2 m (E_{n} -V(x))}
\end{equation*} be the momentum of a classical particle with mass $m$ and
energy $E_{n}$ located at  $x$.  Let
$a_{n-}<a_{n+}$ be the turning points of the classical motion, and let 
\begin{equation*}
  \tau_{n} (x) = \int_{a_{n-}}^{x} dx' \, \frac{m}{p_{n}(x')}
\end{equation*} 
be the time which the particle would take classically to
get from 
$a_{n-}$ to $x$.  Let
\begin{equation*}
  T_{n} = 2 \tau_{n} (a_{n+})
\end{equation*}
be the classical period.  Define 
\begin{equation*}
  S_{n} (x) = \int_{a_{n-}}^{x} dx' \, p_{n}(x') + \frac{h}{8}
\end{equation*} 
Provided that $x$ is not close to one of the classical
turning points we then have
\begin{equation*}
  \overlap{x}{n}
\approx 
      2 \left( \frac{m}{T_{n} p_{n}(x)}\right)^{\frac{1}{2}} \Theta_{n}
(x) 
      \sin \left( \frac{S_{n} (x)}{\hbar}\right) 
\end{equation*}  
where
\begin{equation*}
  \Theta_{n} (x) =\begin{cases}
      1 & \text{if} \quad a_{n-} < x <a_{n+} \\
     0  & \text{if} \quad x < a_{n-} \quad \text{or} \quad  a_{n+}<x
  \end{cases}
\end{equation*}
We are interested in the case when the system is in a
state of the form defined by Eq.~(\ref{eq:  ApproxEState}).  At time $t$
we have (in the Schr\"{o}dinger picture)
\begin{equation*}
   \ket{\psi_{t}} =  \sum_{r=-\frac{\Delta n}{2}}^{r=\frac{\Delta n}{2}}
c_{r} \exp\left(-\frac{i E_{\bar{n}+r} t}{\hbar}\right) \ket{\bar{n} +r}
\end{equation*} 
and
\begin{align}
  \overlap{x}{\psi_{t}} & \approx \sum_{r=-\frac{\Delta
n}{2}}^{r=\frac{\Delta n}{2}}
  i c_{r} \left(\frac{m}{T_{\bar{n}+r}
p_{\bar{n}+r}(x)}\right)^{\frac{1}{2}}
          \Theta_{\bar{n}+r} (x)
\notag
\\ & \hspace{0.5 in} \times
         \left( \exp\left[-\frac{i}{\hbar}\bigl(S_{\bar{n}+r}(x) +
E_{\bar{n}+r} t\bigr)
\right]-\exp\left[\frac{i}{\hbar}\bigl(S_{\bar{n}+r}(x) - E_{\bar{n}+r}
t\bigr) \right] 
\right)
\label{eq:  WKBwvfnc}
\end{align} 
Since we are assuming that $\Delta n \ll \bar{n}$ we can make
some further approximations.  Define
\begin{align*} 
p(x) &=p_{\bar{n}}(x) & E & = E_{\bar{n}} & S(x) & =
S_{\bar{n}} (x) \\
\tau(x) & =\tau_{\bar{n}}(x) & T & = T_{\bar{n}} & a_{\pm} & = a_{\bar{n}
\pm}
\end{align*} 
If $n-\bar{n} \ll \bar{n}$ we can approximate
\begin{equation*}
  p_{n}(x) - p(x) \approx \frac{m}{\sqrt{2 m
(E-V(x))}}(E_{n}-E)=\frac{m}{p(x)}(E_{n}-E)
\end{equation*}
We then use the quantisation condition
\begin{equation*}
  \int_{a_{n-}}^{a_{n+}} dx \, p_{n} (x) = (2 n+1) \frac{h}{4}
\end{equation*} 
to deduce
\begin{align*}
  E_{n} & \approx E+ (n-\bar{n}) \hbar \omega 
\\
\intertext{and}
 S_{n}(x) & \approx S(x) + (n-\bar{n}) \hbar \omega \tau(x)
\end{align*} 
where $\omega$ is the classical frequency, $ 2 \pi/T$.
Using these approximations in Eq.~(\ref{eq:  WKBwvfnc}) we find
\begin{equation}
   \overlap{x}{\psi_t}
\approx
   i \left( \exp \left[ -\frac{i}{\hbar} \bigl(S(x)+ E t\bigr)\right]
g_{-}(x,t)-
            \exp \left[ \frac{i}{\hbar} \bigl(S(x)- E t\bigr)\right]
g_{+}(x,t) \right)
\label{eq:  proxWKBwvfnc}
\end{equation} 
(except in the vicinity of the classical turning points). 
In this expression we have set
\begin{align}
  g_{\pm} (x,t) & = \left(\frac{m}{T p(x)}\right)^{\frac{1}{2}}
                  \sum_{r=-\frac{\Delta n}{2}}^{\frac{\Delta n}{2}}
                  c_{r} \exp \left[\pm i r \omega \bigl(\tau(x) \mp
t\bigr)\right] \Theta (x)
\label{eq:  gplmin} 
\\
\intertext{where}
\Theta (x) = \Theta_{\bar{n}} (x)
\notag
\end{align} 
The imaginary exponentials $\exp\left[ \pm i
\bigl(S(x)\mp E t\bigr)/\hbar\right]$ are rapidly oscillating functions of $x$
and $t$, having spatial period
$h/p(x)$  and  frequency
$E/\hbar$.   The functions $g_{\pm}$, by contrast, are much more
slowly varying, being effectively constant over distances $\ll
(a_{+}-a_{-})/\Delta n$ and times
$\ll T/\Delta n$.

From the form of the expression on the right hand side of 
Eq.~(\ref{eq:  gplmin}) it can be seen that the function $g_{+}(x,t)$
propagates to the right at the classical speed
$p(x)/m$ until it reaches the point $x=a_{+}$, where it is
reflected and becomes the function
$g_{-}(x,t)$.  Similarly, $g_{-}(x,t)$ propagates to the left at the
classical speed until it reaches the point $x=a_{-}$, where it is
reflected and becomes the function $g_{+}(x,t)$.

Let us now calculate the instantaneous Bohmian velocity, given by 
\begin{equation}
  \vb(x,t) = \frac{\hbar \imaginary \left(\overlap{\psi_t}{x}
\frac{\partial}{\partial x}
\overlap{x}{\psi_t} \right)}{m \left|\overlap{x}{\psi_t}\right|^2}
\label{eq:  BohmVelDef}
\end{equation} 
We have from Eq.~(\ref{eq:  proxWKBwvfnc})
\begin{multline*}
  \frac{\partial}{\partial x}
  \overlap{x}{\psi_t}
\approx \Biggl( \exp \left[-\frac{i}{\hbar} \bigl(S(x)+E t\bigr)\right] 
 \left(\frac{p(x)}{\hbar} g_{-} (x,t) + i \frac{\partial}{\partial x}
g_{-}(x,t)\right)\Biggr.
\\
 + \Biggl.
 \exp \left[\frac{i}{\hbar} \bigl(S(x)-E t\bigr)\right]
 \left(\frac{p(x)}{\hbar} g_{+} (x,t) - i \frac{\partial}{\partial x}
g_{+}(x,t)\right)\Biggr)
  \Theta(x)
\end{multline*} 
(except in the vicinity of the classical turning points). 
The functions 
$g_{\pm} (x,t)$ are effectively constant over distances $\sim$ the de
Broglie wavelength.  We may therefore approximate
\begin{equation}
\frac{p(x)}{\hbar} g_{\pm} (x,t) \mp i \frac{\partial}{\partial x}
g_{\pm}(x,t)
\approx \frac{p(x)}{\hbar} g_{\pm} (x,t)
\label{eq:  ProxForVelFormula}
\end{equation} 
It is convenient to write $g_{\pm}$ in modulus-argument
form:
\begin{equation}
  g_{\pm}(x,t) = \sqrt{\rho_{\pm}(x,t)} e^{i \phi_{\pm}(x,t)}
\label{eq:  RhoPhiDef} 
\end{equation} In terms of these quantities, and using the approximation of
Eq.~(\ref{eq:  ProxForVelFormula}), we  have
\begin{equation}
 \imaginary \left(\overlap{\psi_t}{x} \frac{\partial}{\partial x}
  \overlap{x}{\psi_t} \right)
\approx
     \frac{p(x)}{\hbar} \bigl( \rho_{+} (x,t) - \rho_{-} (x,t)\bigr)
\label{eq:  BohmVelNumerator}
\end{equation} 
and
\begin{multline}
  \left| \overlap{x}{\psi_t}\right|^2
\approx
  \rho_{+} (x,t) + \rho_{-} (x,t)
\\
  - 2 \sqrt{\rho_{+} (x,t)  \rho_{-} (x,t)}
    \cos \left(\frac{2 S(x)}{\hbar} +\phi_{+}(x,t) -\phi_{-} (x,t)\right)
\label{eq:  xProbDensity}
\end{multline} 
As $x$ varies the last term on the right hand side of
Eq.~(\ref{eq:  xProbDensity}) fluctuates rapidly, with a  spatial period
$\sim$ the de Broglie wavelength.  The functions
$\rho_{\pm}$, by contrast, are nearly constant on this scale.  It follows
that the quantity
\begin{equation}
 \bar{\rho}(x,t) = \rho_{+}(x,t) + \rho_{-} (x,t)
\label{eq:  xMeanProbDensity}
\end{equation} 
is the mean $x$-space probability density function,
averaged over a de Broglie wavelength.

Inserting the results of Eqs.~(\ref{eq:  BohmVelNumerator}) 
and~(\ref{eq:  xProbDensity}) in Eq.~(\ref{eq:  BohmVelDef}) we find
\begin{multline}
 \vb(x,t) \approx 
  v_{\mathrm{cl}}(x) \\ \times
    \frac{\rho_{+}(x,t)-\rho_{-}(x,t)
        }{\rho_{+}(x,t)+\rho_{-}(x,t)-
             2\sqrt{\rho_{+}(x,t)\rho_{-}(x,t)} 
             \cos \left(\frac{2}{\hbar}
S(x)+\phi_{+}(x,t)-\phi_{-}(x,t)\right)
          }
\label{eq:  BohmVelFinal}
\end{multline} 
for $a_{-}<x<a_{+}$ (except in the immediate vicinity of
the turning points).   In this expression
$v_{\mathrm{cl}}(x)=p(x)/m$, the classical speed at position $x$.

$\rho_{\pm}$, $\phi_{\pm}$ are slowly varying functions of $x$.  On the other
hand the term
$2 S(x)/\hbar$ is very rapidly varying.  It follows that $\vb(x,t)$ varies
rapidly between the  extremal values 
\begin{equation*}
  v_{\pm}(x,t) = v_{\mathrm{cl}}(x)
\frac{\sqrt{\rho_{+}(x,t)}\pm\sqrt{\rho_{-}(x,t)}
    }{\sqrt{\rho_{+}(x,t)}\mp\sqrt{\rho_{-}(x,t)}}
\end{equation*} 
over distances $\sim$ the de Broglie wavelength.

If $\rho_{+}(x,t)\gg\rho_{-}(x,t)$ 
\begin{equation*}
  v_{-}(x,t)\approx v_{+}(x,t)\approx v_{\mathrm{cl}}(x)
\end{equation*} 
and the motion is approximately classical.  If, on the
other hand,
$\rho_{+}(x,t)\ll\rho_{-}(x,t)$
 \begin{equation*}
  v_{-}(x,t)\approx v_{+}(x,t)\approx - v_{\mathrm{cl}}(x)
\end{equation*} 
The motion is again approximately classical, but in the
opposite direction.

Suppose, however, that neither of these conditions is satisfied. In that
case 
$|v_{+}(x,t)|\gg v_{\mathrm{cl}}(x)$, and the motion is highly
non-classical.

The necessary and sufficient condition for the Bohmian velocity to be
close to the one of the two possible values of the classical velocity at
position $x$ is, therefore,
\begin{equation}
  \frac{1}{2}\left(
    \frac{\rho_{+}(x,t)}{\rho_{-}(x,t)}+\frac{\rho_{-}(x,t)}{\rho_{+}(x,t)}
        \right)
\gg 1
\label{eq:  classCond}
\end{equation} 
(except in the vicinity of the points $x=a_{\pm}$).

If we only require the state to be such that there is a \emph{high
probability} of the  Bohmian velocity being close to $\pm
v_{\mathrm{cl}}(x)$ at all times, then we only need to impose
condition~(\ref{eq:  classCond}) at points where the mean probability
density
$\bar{\rho}=\rho_{+}+\rho_{-}$ is non-negligible [see the remark following 
Eq.~(\ref{eq:  xMeanProbDensity})].  It is, however, important that the
inequality always holds true at such points, for every time $t$.  Suppose,
for example, that at a particular instant the functions $\rho_{\pm}$ are
as shown in  Fig.~\ref{fig: rho}(a).  
\begin{figure}[htb]
\begin{picture}(460,340)
%\put(0,0){\framebox(460,340)}
\put(0,210){\includegraphics{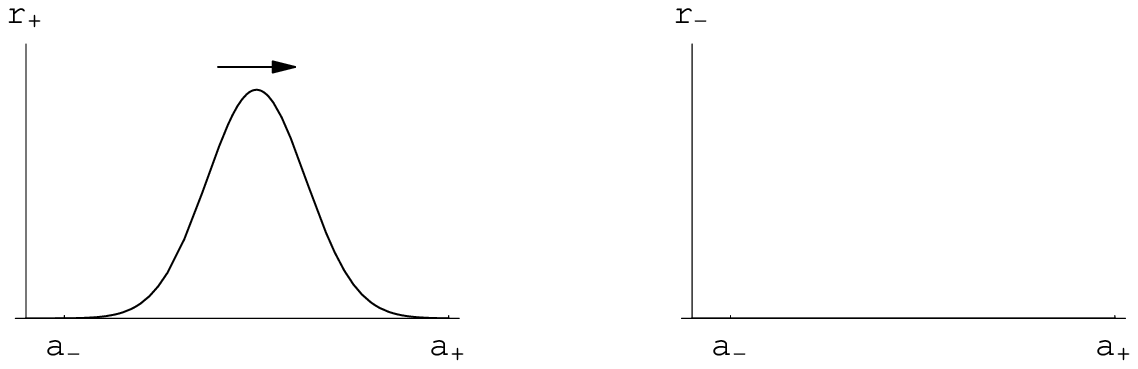}}
\put(192,196){\makebox{(a)}}
\put(0,40){\includegraphics{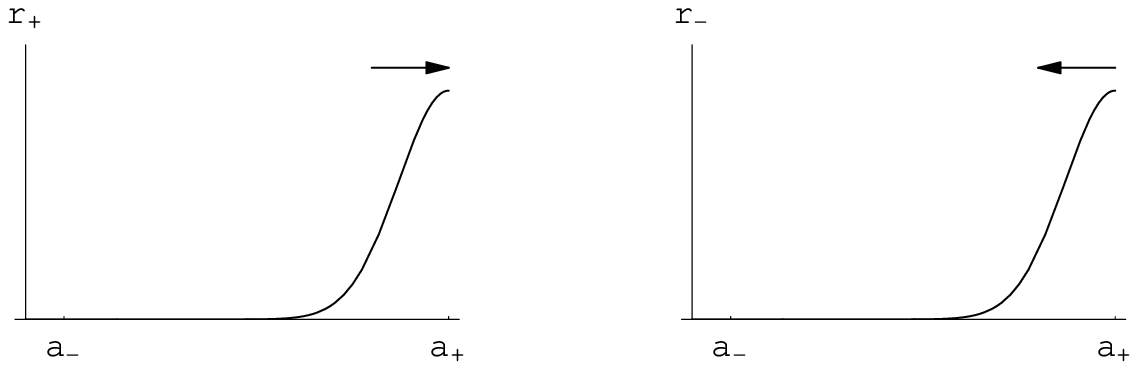}}
\put(192,27){\makebox{(b)}}
\end{picture}
\caption{The graphs in (a) show the functions $\rho_{\pm}$ at a time when
$\rho_{+}$ has a single, rather broad peak, centred in the middle of the
interval
$(a_{-},a_{+})$, and when $\rho_{-}$ is everywhere negligible.  In this
situation  $\vb$ is everywhere close to $v_{\mathrm{cl}}$.  The graphs in
(b) show the situation at a time
$\sim \frac{T}{4}$ later.  There is then a significant probability of the
particle being in a region where $\vb$ fluctuates violently, up to a
maximum which is much greater than the speed which would be expected
classically.  The arrows show the direction of propagation of the
functions.  See the discussion in the paragraph following 
Eq.~(\ref{eq:  classCond})}
\label{fig:  rho}
\end{figure} 
Then  $\vb \approx v_{\mathrm{cl}}$ at all values of
$x$ for which
$\bar{\rho}$ is non-negligible.  However, at a time $\sim T/4$
later $\rho_{\pm}$  will be as shown in  Fig.~\ref{fig: rho}(b), so that
there is a signficant probability of the particle being in a region where
$\rho_{+} \approx \rho_{-}$, and which is well away from the turning
points.   From a consideration of this and other examples it can be seen
that there will only be a high probability of the velocity being close to
$\pm v_{\mathrm{cl}}$ throughout the motion if the state is a
highly localised wave packet, so that the peak in 
$\bar{\rho}$ is very narrow.

Of course, there will be times when  $\rho_{+} \approx
\rho_{-}$ at the turning points, even when the peak in
$\bar{\rho}$ is very narrow.  However, this does not invalidate the
conclusion.  In the first place, the
approximations leading to Eq.~(\ref{eq:  BohmVelFinal}) break down when
$x\approx a_{\pm}$.  In the second place, even if Eq.~(\ref{eq: 
BohmVelFinal}) were valid at the turning points, the fact that
$v_{\mathrm{cl}}\approx 0$ at these points means that one can still have
$v_{\pm}
\approx v_{\mathrm{cl}}$, even though $\rho_{+}\approx \rho_{-}$

\section{The Time-Averaged Velocity}
\label{sec:  TimeAve} 
We saw in the last section  that the instantaneous Bohmian
velocity is typically very rapidly fluctuating.  However, classical
physics is based on observations, not of the instantaneous velocity, but
rather of the velocity averaged over a finite time interval. 
Furthermore, the averaging time may be assumed to be large in comparison
with the time-scale of the fluctuations in the instantaneous Bohmian
velocity.  In order to make the argument complete we consequently need to
consider the possibility~\cite{Squires} that the
time-averaged Bohmian velocity may be consistent with the predictions of
classical physics, even when the
instantaneous velocity is not.

Referring to Eq.~(\ref{eq:  BohmVelFinal}) we see that the equation of
motion is, approximately,
\begin{equation*}
  \frac{d x}{d t}
\approx  
   \frac{v_{\mathrm{cl}}(x) \bigl(\rho_{+}(x,t)-\rho_{-}(x,t)\bigr) }{
    \rho_{+}(x,t)+\rho_{-}(x,t)-2\sqrt{\rho_{+}(x,t) \rho_{-}(x,t)}
              \cos \left[\frac{2}{\hbar} S(x)
+\phi_{+}(x,t)-\phi_{-}(x,t)\right]}
\end{equation*} 
We wish to solve this equation subject to the initial condition $x=x_0$
when 
$t=t_0$.  We may assume
$(t-t_{0})\ll
T/\Delta n$  and $|x-x_0|\ll
(a_{+}-a_{-})/\Delta n$.  We can then further approximate
\begin{equation*}
  \frac{d x}{d t}
\approx
  \frac{ v_{\mathrm{cl}}(x_0) \sinh \chi_0}{\cosh \chi_0 -\cos \bigl[
\phi_{0} + \frac{4
\pi}{\lambda_{0}}(x-x_{0})\bigr]}
\end{equation*} 
where
\begin{align*}
   \chi_0 & = \frac{1}{2} \log \left(
\frac{\rho_{+}(x_0,t_0)}{\rho_{-}(x_0,t_0)}\right)
\\
  \phi_0 & = \frac{2}{\hbar} S(x_0) +\phi_{+}(x_0,t_0) - \phi_{-}(x_0,t_0)
\end{align*} 
and where $\lambda_{0}=h/p(x_0)$ is the
de Broglie wavelength at position
$x_0$.  The solution to this equation is
\begin{multline}
  \left( 1 - \frac{\lambda_{0} \sech \chi_0}{4 \pi (x-x_0)}
          \left[ \sin \left(\phi_0 +\tfrac{4 \pi}{\lambda_{0}}
(x-x_0)\right) -\sin\phi_0\right]\right)(x-x_0)
\\
\approx \left( v_{\mathrm{cl}}(x_0) \tanh \chi_0\right) (t-t_0)
\label{eq:  trajectory}
\end{multline} 
If $x-x_0 \gg \lambda_0$ we may write
\begin{equation*}
  x \approx x_0 + \bvb (t-t_0)
\end{equation*} 
where $\bvb$ is the time-averaged velocity
\begin{equation}
  \bvb = \frac{1}{\tau}
         \int_{t_0}^{t_0+\tau} dt \, \frac{d x}{d t} 
       \approx  v_{\mathrm{cl}}(x_0) \tanh \chi_0
\label{eq:  VTResult}
\end{equation}
$\tau$ being the time to move one de Broglie wavelength, 
$\lambda_0/\bigl(v_{\mathrm{cl}}(x_0)\tanh\chi_0\bigr)$.

\begin{figure}[htb]
\begin{picture}(460,250)
%\put(0,0){\framebox(460,250)}
\put(0,0){\includegraphics{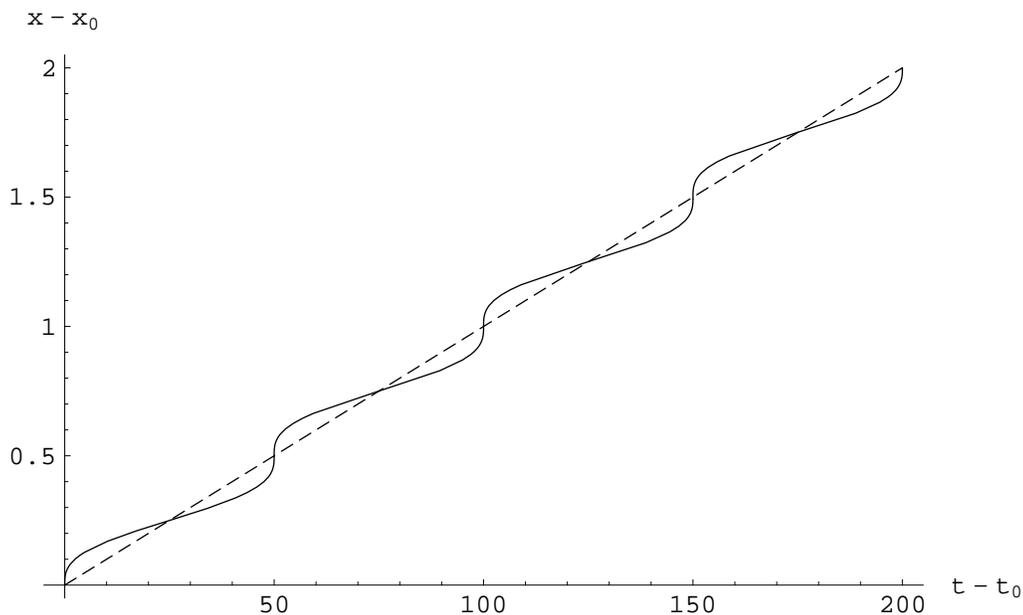}}
\end{picture}
\caption{The solid line shows the dependence of $x$ on $t$ for the case
$\phi_0=0$,
$\chi_0=0.01$.  See Eq.~(\ref{eq:  trajectory}).  The broken line shows
the time-averaged trajectory $x =x_0+\bvb (t-t_0)$.   Units have been
chosen so that $\lambda_0 =v_{\mathrm{cl}}(x_0)=1$.   }
\label{fig:  traject}
\end{figure}

\begin{figure}[p]
\begin{picture}(460,470)
%\put(0,0){\framebox(460,470)}
\put(0,240){\includegraphics{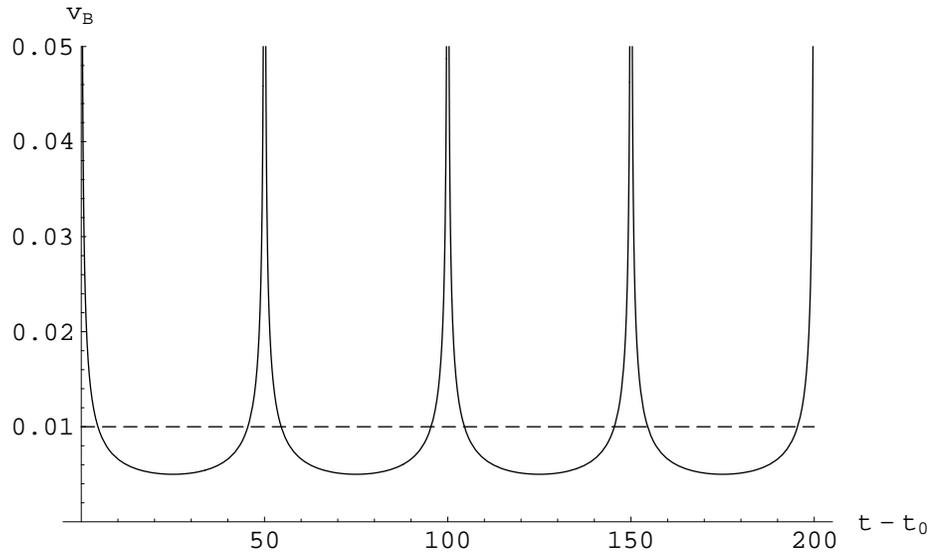}}
\put(190,220){\makebox{(a)}}
\put(0,20){\includegraphics{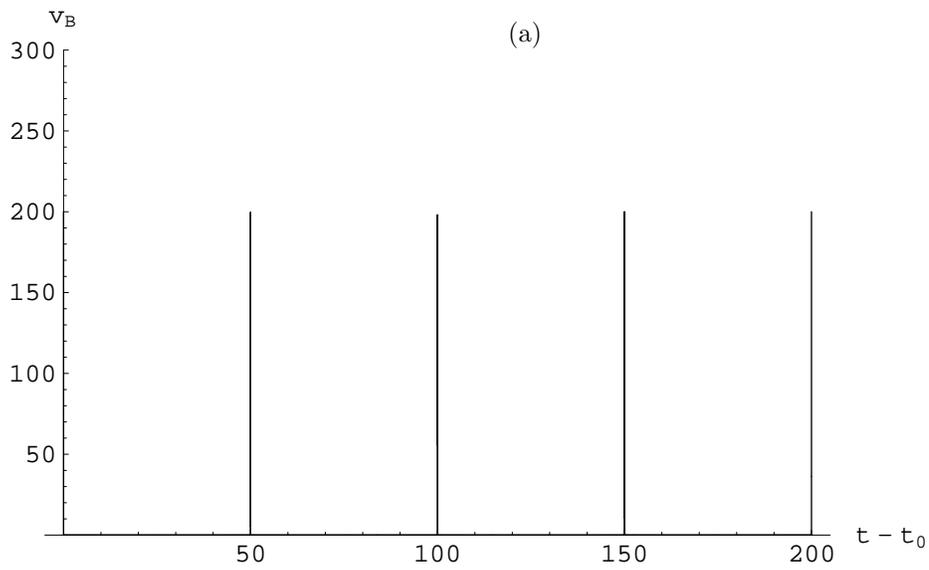}}
\put(190,0){\makebox{(b)}}
\end{picture}
\caption{In graph (a) the solid line shows the dependence of velocity on
time for the case
$\phi_0=0$,
$\chi_0=0.01$, with units chosen as in Fig.~\ref{fig:  traject}, so that
$\lambda_0 =v_{\mathrm{cl}}(x_0)=1$.  The broken line shows the
time-averaged velocity $\bvb$ ($=0.01$
$\times$ the classical velocity).  It can be seen that during the greater
part of the motion the particle is travelling more slowly than this. In
(b) the graph is reproduced with a different choice of scale on the $v$
axis, so as to include the  maxima at
$200$ $\times$ the classical velocity.  It can be seen that the peaks are
extremely narrow.}
\label{fig:  velocity}
\end{figure}

In  Fig.~\ref{fig:  traject}
we illustrate this result by plotting $x-x_0$ as a function of
$t-t_0$ for the case
$\phi_0 = 0$,
$\chi_0 = 0.01$ (implying $\frac{\rho_{+}}{\rho_{-}}=1.02$ and $\bvb=0.01
v_{\mathrm{cl}}(x_0)$).  The behaviour  of the velocity as a function of
time is illustrated in  Fig.~\ref{fig:  velocity}.

The graphs in Fig.~\ref{fig:  velocity} illustrate the fact that if
$\chi_{0}\approx 0$ (so that
$\rho_{+}\approx\rho_{-}$) the speed  peaks at a value which is very much
greater than $v_{\mathrm{cl}}$.  However, they also show that this
phenomenon is of very short duration.  An equally significant feature of
the motion is the fact that the  speed is, for the most part, very much
smaller than $v_{\mathrm{cl}}$.  As a result $|\bvb| \ll
v_{\mathrm{cl}}$.

The condition for  $\bvb$ to be close to 
$\pm v_{\mathrm{cl}}$ is that $|\chi_0|\gg 0$.  This is the same as  the
condition derived in the last section [\emph{c.f.} 
Eq.~(\ref{eq:  classCond})].  Consequently, the conclusion still stands,
that there is only a high probability of the motion being quasi-classical
in the case of a narrowly localised
wave-packet.
\section{Comparison with Garc\'{i}a de Polavieja's Interpretation}
\label{sec:  GdeP}
We now consider  the underlying reasons for the 
behaviour discussed in the last two sections.
It might be thought that the counter-intuitive behaviour of the
Bohmian trajectories (of an isolated particle) is not very surprising. 
According to the Copenhagen school of thought the concept of a precisely
defined, completely determinate trajectory is  illegitimate.  The
Copenhagen interpretation is not so widely accepted as once was the case. 
Nevertheless, there continues to be a widespread feeling that the concept
of a determinate trajectory, though  not excluded in point of
strict logic, represents an artificial construction which is imposed on
the theory by arbitrary
\emph{fiat}.  Someone who takes this point of view may feel  that it is
only to be expected that the trajectories will tend to be
strikingly counter-intuitive.

In fact, however, it would appear from the work of Garc\'{i}a de
Polavieja~\cite{Polavieja} that there is at least one  interpretation of
this kind in which the trajectories are much better behaved, and in
which one already obtains the correct classical limit even in the case of
an isolated system, without having to take into account the effect of
the environment.  This suggests that the counter-intuitive behaviour of
the Bohmian trajectories may actually be due, not to constraints inherent
in the very idea of a trajectory interpretation, but rather to features
specific to the particular manner in which the Bohm interpretation
realises this idea.  We will now try to identify these features, by
making a comparison between the Bohm interpretation and Garc\'{i}a de
Polavieja's interpretation.

In  interpretations of the kind we are considering one has to make a
choice as to the intrinsic probability distribution describing the
``beables'' of the theory.  In the Bohm interpretation one starts with an
intrinsic configuration space probability 
distribution, which is taken to be $|\overlap{x}{\psi}|^2$. 
In Garc\'{i}a de Polavieja's interpretation, by contrast, one starts with
an intrinsic phase space probability distribution, which is taken to be
the Husimi,  or
$Q$-function defined by~\cite{Hus,Hil,Lee,Leon,Torres}
\begin{equation}
  Q_{\lambda}(x,p) = \frac{1}{h} \left|
\boverlap{\coh{x}{p}{\lambda}}{\psi}\right|^2
\label{eq:  HusTermsCoh}
\end{equation} 
where $\ket{\coh{x}{p}{\lambda}}$ is the coherent state
with $x$-representation wave function
\begin{equation}
  \boverlap{x'}{\coh{x}{p}{\lambda}} = \left(\frac{1}{\pi
\lambda^2}\right)^{\frac{1}{4}}
  \exp \left[ - \frac{1}{2 \lambda^2} (x'-x)^2 + \frac{i}{\hbar} p x' -
\frac{i}{2 \hbar} p x
  \right]
\label{eq:  CohSteDef}
\end{equation} 

The significance of the parameter
$\lambda$ can be understsood by referring to the role that
$Q_{\lambda}(x,p)$ plays in the theory of measurement. The function
$|\overlap{x}{\psi}|^2$ gives the distribution of results for ideal,
perfectly accurate measurements of position only.  
$Q_{\lambda}(x,p)$, by contrast, gives the distribution when one makes
simultaneous, imperfectly accurate measurements of both position
and momentum~\cite{Arthurs,Ali,Self3} (for  reviews of the theory
of simultaneous measurement processes, and additional references, see
Busch~\cite{Busch} and Leonhardt~\cite{Leon}).  The significance of the
parameter
$\lambda$ is that it specifies the relative accuracy of the measurements
of $x$ and
$p$.  That is, $Q_{\lambda}(x,p)$ describes the outcome when
$x$ is measured to retrodictive accuracy $\RErr x =\pm
\lambda/\sqrt{2}$ and
$p$ is measured to retrodictive accuracy $\RErr p =\pm
\hbar/(\sqrt{2}\lambda)$, and when, in addition, the measurements
of $x$ and $p$  are retrodictively unbiased~\cite{Self3}.  Such
measurements are optimal, in the sense that the lower bound set by the 
inequality\footnote{As explained in ref.~\cite{Self4} this inequality
is not the same as the uncertainty principle usually so called. The
quantities
$\RErr x$,
$\RErr p$ are errors, not uncertainties.}
$\RErr x \RErr p \ge \hbar/2$ is actually achieved~\cite{Self4}.

It can be seen that Garc\'{i}a de Polavieja's interpretation is in fact,
not a single interpretation, but rather an infinite family of
interpretations, parameterised by
$\lambda$.  In order to obtain  the correct classical limit for an
isolated system the value of $\lambda$ must be appropriately chosen.  To
see this, let us calculate 
$Q_{\lambda}$ for states of the type defined by Eq.~(\ref{eq: 
ApproxEState}). 

Eqs.~(\ref{eq:  proxWKBwvfnc}) and~(\ref{eq: 
CohSteDef}) imply
\begin{multline}
\boverlap{\coh{x}{p}{\lambda}}{\psi_t}
\\
\approx i \left(\frac{1}{\pi \lambda^2}\right)^{\frac{1}{4}}
\exp \left[-\frac{i}{\hbar}\Bigl(E t - \frac{1}{2} p x\Bigr)\right]
\int dx' \, \exp\left[-\frac{1}{2 \lambda^2} (x'-x)^2 - \frac{i}{\hbar} p
x'
                 \right]
\\ \times
              \Biggl(  \exp \left[ - \frac{i}{\hbar} S(x')  \right] g_{-}
(x')
                     - \exp \left[ \frac{i}{\hbar} S(x')  \right] g_{+}
(x')
              \Biggr)
\label{eq:  HusCalcA}
\end{multline} 
away from the classical turning points.
Suppose that
\begin{equation}
 \lambda \ll  \lambda_{+}(x)=\min \left(\frac{a_{+}-a_{-}}{\Delta n},
   \left( \frac{\hbar}{|p'(x)|}\right)^{\frac{1}{2}}\right)
\label{eq:  LambdaPlusDef} 
\end{equation} 
We can then approximate
\begin{equation*}
g_{\pm}(x') \approx g_{\pm}(x)
\hspace{0.5 in} \text{and} \hspace{0.5 in}
S(x') \approx S(x) + p(x) (x'-x)
\end{equation*}
Making these approximations in Eq.~(\ref{eq:  HusCalcA}), carrying out
the  Gaussian integration, and substituting the result 
in~(\ref{eq:  HusTermsCoh}) gives
\begin{multline} Q_{\lambda} (x,p)
\approx
\frac{\lambda}{\sqrt{\pi} \hbar}
\Biggl\{ \exp\left[ -\frac{\lambda^2}{\hbar^2} \bigl(
p+p(x)\bigr)^2\right]
                      \rho_{-}(x)
       + \exp\left[ -\frac{\lambda^2}{\hbar^2} \bigl(
p-p(x)\bigr)^2\right]
                      \rho_{+}(x)
\Biggr.
\\
\Biggl.
      - 2 \exp\left[ -\frac{\lambda^2}{\hbar^2} 
                             \left( p^2 + \bigl(p (x)\bigr)^2\right)
              \right]
         \cos\left[ \frac{2}{\hbar} S(x) + \phi_{+}(x) -\phi_{-} (x)
\right] \sqrt{\rho_{+}(x) \rho_{-}(x)}
\Biggr\}
\label{eq:  HusCalcB}
\end{multline} 
where $\rho_{\pm}$, $\phi_{\pm}$ are the quantities defined 
by Eq.~(\ref{eq:  RhoPhiDef}).
Suppose that we also have 
\begin{equation}
\lambda \gg 
\lambda_{-}(x)=\frac{\hbar}{p(x)}
\label{eq:  LambdaMinusDef}
\end{equation}
In that case the third, oscillatory term in
parentheses on the right hand side of Eq.~(\ref{eq:  HusCalcB}) is
negligible.  Also, the Gaussian peaks in the first and second terms are
very narrow in comparison with the classical momentum $p(x)$,
and may therefore be regarded as approximate
$\delta$-functions.  We conclude
\begin{equation} 
Q_{\lambda} (x,p)
\approx
\delta\bigl(p+p(x)\bigr) \rho_{-} (x) +
\delta\bigl(p-p(x)\bigr) \rho_{+} (x)
\label{eq:  HusClass}
\end{equation} 
provided that $\lambda_{-}(x) \ll \lambda \ll \lambda_{+}(x)$.
This is a possible classical distribution for a particle of energy
$E_{\bar{n}}$---which suggests that the trajectories in Garc\'{i}a de
Polavieja's interpretation will also be approximately classical when 
$\lambda$ lies within the stated range.

Suppose, on the other hand, that $\lambda \ll \lambda_{-}(x)$. In that
case the widths of the Gaussian peaks on the right hand side of
Eq.~(\ref{eq:  HusCalcB}) are much larger than
$p(x)$, so that we have, approximately, 
\begin{multline*}
  Q_{\lambda}(x,p)
\approx
  \frac{\lambda}{\sqrt{\pi} \hbar}
  \exp\left[ - \frac{\lambda^2}{\hbar^2} p^2\right]
\\ \times
  \Biggl\{\rho_{-}(x)+\rho_{+} (x) - 2 \sqrt{\rho_{-}(x) \rho_{+}(x)}
         \cos\left[\frac{2}{\hbar} S(x) +\phi_{+}(x) - \phi_{-}(x)\right]
     \Biggr\}
\end{multline*} 
Comparing this expression with Eq.~(\ref{eq:  xProbDensity}) we see that 
\begin{equation}
  Q_{\lambda}(x,p)
\approx
  \frac{\lambda}{\sqrt{\pi} \hbar}
  \exp\left[ - \frac{\lambda^2}{\hbar^2} p^2\right]
  \bigl| \overlap{x}{\psi_t}\bigr|^2
\label{eq:  HusCalcC}
\end{equation} 
If, on the other hand, $\lambda \rightarrow \infty$, then it is not
difficult to show~\cite{Self5}
\begin{equation}
  Q_{\lambda}(x,p)
\approx
  \frac{1}{\sqrt{\pi} \lambda}
  \exp\left[ - \frac{1}{\lambda^2} x^2\right]
  \bigl| \overlap{p}{\psi_t}\bigr|^2
\label{eq:  HusCalcD}
\end{equation} 
These distributions are both highly non-classical.  It follows that the
trajectories in Garc\'{i}a de Polavieja's interpretation will also be
highly non-classical in the limit as $\lambda$ becomes very small, or
very large.

One can understand the reason why $Q_{\lambda}$ behaves in this way if
one considers its interpretation as the probability distribution
describing the outcome of a retrodictively optimal joint measurement of
$x$ and
$p$~\cite{Self3}. Eq.~(\ref{eq:  HusCalcC}) describes a situation in
which the error in $x$ is very small, and the error in  $p$
is correspondingly large.  The fact that the error in $p$ is 
large means that the measurement is too insensitive to pick up any
correlation between the values of $x$ and $p$.  On the other hand the
fact that the error in $x$ is small means that the measurement is able
pick up the very rapid variation in the probability distribution which
occurs in the direction parallel to the $x$ axis, over distances 
$\sim$ the de Broglie wavelength.  Analogous statements apply to 
Eq.~(\ref{eq:  HusCalcD}), except that now it is the measurement of $p$
which is very accurate, and the measurement of $x$ which is
correspondingly inaccurate.
Eq.~(\ref{eq:  HusClass}), by contrast, describes a situation in which
both $x$ and $p$ are measured to an intermediate degree of accuracy. The
errors are both  small enough to ensure that the measured values of
$x$ and $p$ are highly correlated.  At the same time, they are
both large enough to ensure that the measurement is insensitive
to the very rapid variations in the functions $|\overlap{x}{\psi_{t}}|^2$
and
$|\overlap{p}{\psi_{t}}|^2$.

Classical physics is based on situations in which  $x$ and $p$ have both
been determined to an intermediate degree of accuracy---which is the why
the distribution of Eq.~(\ref{eq:  HusClass}) is of  classical form.

Let us now relate this discussion to the behaviour of the Bohmian
trajectories. Let $\bvl (x)$ be the mean velocity at $x$ in Garc\'{i}a de
Polavieja's interpretation:
\begin{equation*}
  \bvl (x) =
  \frac{ \int dp\, p Q_{\lambda} (x,p)}{m \int dp \, Q_{\lambda}(x,p)}
\end{equation*}
$\bvl(x)$ is also the mean velocity which would be observed at $x$ if one
were to make retrodictively optimal joint measurements of $x$ and $p$ to
accuracies $\pm \lambda/\sqrt{2}$ and
$\pm \hbar/(\sqrt{2} \lambda)$ respectively.  Substituting the
expression given by Eq.~(\ref{eq:  HusCalcB}) in this equation and taking
the limit as
$\lambda \rightarrow 0$ gives
\begin{equation*}
  \lim_{\lambda \rightarrow 0}  \bigl( \bvl (x) \bigr)
=
  v_{\mathrm{cl}}(x)
  \frac{\rho_{+}(x)-\rho_{-}(x)
      }{\rho_{+}(x)+\rho_{-}(x)
           - 2 \sqrt{\rho_{+}(x) \rho_{-}(x)} 
        \cos\left[  \frac{2}{\hbar} S(x) +\phi_{+}(x)-\phi_{-}(x)
                           \right]
      }
\end{equation*}
where $v_{\mathrm{cl}}(x)=p(x)/m$.
Comparing with Eq.~(\ref{eq:  BohmVelFinal}) we see that
\begin{equation*}
  \lim_{\lambda \rightarrow 0}  \bigl( \bvl (x) \bigr)
= \vb (x)
\end{equation*}
In Section~\ref{sec:  WKB}  we saw that 
the instantaneous Bohmian speed $|\vb(x)|$ tends to take values which are
much larger than the  classical value, while in Section~\ref{sec: 
TimeAve} we saw that
$\left|\bvb(x)\right|$ exhibits the opposite behaviour, often taking
values which are much less than the classical speed.  The result just
derived explains both these features.

The reason for the velocity spikes illustrated in Fig.~\ref{fig: 
velocity}(b) is that $\vb(x)$ is the mean observed velocity at $x$ in the
limit as the measurement of position becomes almost perfectly accurate. 
In order to carry out such a measurement it would be necessary to use a
probe whose momentum was large in comparison with the momentum of the
particle.  Under such conditions violent fluctuations in the observed
velocity are not unexpected.

The reason that $\left|\bvb(x)\right|$ is often much less than the
classical speed is that the Bohmian velocity is related specifically to
the 
\emph{mean} observed velocity at $x$.  Suppose, for example, that the
particle was in an exact energy eigenstate.  In that case
$\rho_{-}(x)=\rho_{+}(x)$,  and $\bvl(x) =0$.  In
Garc\'{i}a de Polavieja's interpretation (for intermediate values of
$\lambda$) this implies the classical picture of an ensemble of particles,
one half of which are moving at the classical speed to the right, while
the other half are moving at the classical speed to the left, with only
the
 \emph{mean} velocity 
being zero.  In the Bohm interpretation, by contrast, it implies the
highly non-classical picture of an ensemble in which each \emph{individual}
particle has velocity zero.  The picture is non-classical because it
takes a quantity having the observational significance of a mean, and
interprets it as a property of individual particles.

This feature of the
Bohm interpretation is related to the fact that the equation of motion  
is first-order in time, so that the velocity is a simple function of
position.  The Bohm interpretation is consequently unable to describe a
situation in which both signs of the velocity occur with non-negligible
probability.
\section{Other Trajectory Interpretations}
\label{sec:  Others}
It should be noted that Garc\'{i}a de Polavieja's interpretation 
has certain drawbacks.    In the first place, the equations of
motion involve an infinite series, whose individual terms are often only
defined in a distributional sense, and whose convergence properties are
unclear.  In fact, as we will show in a subsequent article,  the analytic
properties~\cite{Self5} of the Husimi function can be used to re-write
the equations of motion in a different form, which involves an absolutely
convergent series of holomorphic functions.  Nevertheless, it does not
seem to be possible to avoid the use of an infinite series---which is
clearly undesirable from a calculational point of view.  Another possible
difficulty stems from the fact that the range of admissible values of
$\lambda$ depends on the potential.  It is not entirely clear that there
exists a single value of $\lambda$ which would be admissible for all
physically reasonable choices of potential.  It would  be
interesting to know if there exists some other trajectory
interpretation, which also produces the correct
classical limit for an isolated system, but which does not have the  same
disadvantages as Garc\'{i}a de Polavieja's interpretation.  The discussion
in the last section provides some indications as to the direction one
might  take in such an enquiry.

We would particularly stress the significance of the intrinsic phase space
probability
distribution, describing the ``beables'' of the theory.  This function
is usually chosen so as to have, as one of its marginal distributions, 
either the function $|\overlap{x}{\psi}|^2$, or the function
$|\overlap{p}{\psi}|^2$.  Roy and
Singh, in a very interesting series of papers~\cite{Singh1,Singh2,Singh3},
have proposed an interpretation in which the intrinsic phase space
distribution has
both these functions as its marginals.  The distribution
$Q_{\lambda}(x,p)$, by contrast, has \emph{neither} function as a
marginal.
At first sight it may appear that this makes it an unnatural
choice~\cite{Martin}.   However, the discussion in the last section shows
that it actually has some important advantages.  As we saw, it is just
because the Bohm interpretation does have $|\overlap{x}{\psi}|^2$ as the
intrinsic
$x$-space distribution that it tends to produce the velocity spikes
illustrated in Fig.~\ref{fig:  velocity}.

As we remarked in Section~\ref{sec: GdeP}, the reason that the
distribution in  Eq.~(\ref{eq:  HusClass}) is of classical form is that
it describes the kind of measurement on which classical physics is based,
in which 
$x$ and $p$ are both determined to an intermediate degree of accuracy.
The function $|\overlap{x}{\psi}|^2$, by contrast, describes a measurement
in which $x$ is determined with perfect accuracy, and $p$ is not
determined at all.  Such measurements require experimental conditions
which are very unlike  the conditions of our ordinary experience. 
In particular, they involve a very significant perturbation 
of the momentum of the particle whose position is being measured. 
Consequently, it is perhaps not  surprising that an interpretation based
on the function
$Q_{\lambda}(x,p)$ (with
$\lambda$ appropriately chosen) gives trajectories which are much more
nearly classical than one which is based on the function 
$|\overlap{x}{\psi}|^2$.
\section{Conclusion}
In this paper we have only considered the behaviour of an isolated
system.  Of course, the macroscopic bodies of our ordinary experience
never are isolated (by definition since, if they were isolated, we
would not be able to experience them).  Bohm and Hiley argue that the
effect of the interaction with the environment is to cause the Bohmian
trajectory to become quasi-classical.  In the sequel to this paper we
will give some further arguments in support of their conclusion.

The Bohm interpretation is sometimes seen as being in opposition to the
Copenhagen interpretation, so that one has to take up a position either
for or against.  This does not appear to have been the view of Bohm
himself.  On p.5 of \emph{The Undivided Universe} Bohm and Hiley argue
that ``there should be a kind of dialogue between different
interpretations rather than a struggle to establish the primacy of any
one of them''.  One of
the most interesting features of the Bohm interpretation is the way in
which it serves to illuminate some of Bohr's key concepts from a somewhat
unexpected direction (indeed, the title \emph{The Undivided Universe} does 
itself contain an allusion to one of Bohr's concepts). 
The results we have been discussing
provide  some further illustrations of the connection between the Bohm
interpretation and other approaches to the problem of interpretation:  for
they show that the interaction with the environment plays a crucial role
in the Bohm interpretation just as it does in the decoherent
histories~\cite{GellMann,Omnes} and existential~\cite{Zurek1}
interpretations.  

Our discussion also casts some light on the concept of a ``hidden variable''. 
It would be reasonable to say that
the trajectory is hidden if the system is isolated, but not hidden if the
system is open, so that information about the trajectory is recorded in the
environment.  Since the Bohm interpretation makes statements about the
trajectory of an isolated particle, it is therefore reasonable to describe it
as a ``hidden-variables theory''.

There are many different trajectory interpretations, which all make
different predictions regarding the motion of an isolated particle. 
However, this has no bearing on their empirical acceptability since the
motion of an isolated particle is ``hidden'', and so it cannot be empirically
determined (by definition:  the particle cannot be observed by something
external to itself if it is not interacting with something external to
itself).  In order to be empirically acceptable it is only necessary that the
various interpretations all make the same predictions regarding the motion of a
particle which is not isolated.

This point is somewhat reminiscent of Copenhagen doctrines regarding the
role of the external observer. There is, however, an important
difference since the proponents of the Copenhagen interpretation appeared to
make physical processes depend on the actual presence of such an observer. 
They thereby introduced an unacceptable element of subjectivity into physical
theory.  No such subjectivity is present here.  It is indeed the case that,
if  the system has interacted with the rest of the universe, then there is the
\emph{possibility}  of an external observer using the interaction to acquire
information about the system.  However, the interaction is not
\emph{dependent} on this happening.  On the contrary,  it is a completely
objective process which would occur even if there were no observers.


\begin{thebibliography}{99}
\label{sec:  bibliography}
\bibitem{Bohm1}
  D.~Bohm,
    \emph{Phys.\ Rev.\ }\textbf{85}, 166, 180 (1952).
\bibitem{Bohm2}
 D.~Bohm and B.J.~Hiley,
  \emph{The Undivided Universe} (Routledge, London, 1993).
\bibitem{Holland1}
  P.R.~Holland,
  \emph{The Quantum Theory of Motion} (Cambridge University Press, 
  Cambridge, 1993).
\bibitem{Holland2}
  P.R.~Holland,
  in \emph{Bohmian Mechanics and Quantum Theory:  An Appraisal},
  J.T.~Cushing, A.~Fine and S.~Goldstein, eds.\ (Kluwer, Dordrecht, 1996).
\bibitem{Holland3}
  P.R.~Holland,
   \emph{Found.\ Phys.\ }\textbf{28}, 881 (1998).
\bibitem{Englert}
	B.-G.~Englert, M.O.~Scully, G.~S\"{u}ssmann and H.~Walther,
 \emph{Z.~Naturforsch.}\ \textbf{47a},1175 (1992).
\bibitem{Durr}
 D.~D\"{u}rr, W.~Fusseder, S.~Goldstein and N.~Zanghi,
  \emph{Z.~Naturforsch.}\ \textbf{48a}, 1261 (1993).
\bibitem{Englert2}
	B.-G.~Englert, M.O.~Scully, G.~S\"{u}ssmann and H.~Walther,
 \emph{Z.~Naturforsch.}\ \textbf{48a},1263 (1993).
\bibitem{Dewdney}
 C.~Dewdney, L.~Hardy and E.J.~Squires,
  \emph{Phys.~Lett.~A}\ \textbf{184}, 6 (1993).
\bibitem{Aharanov}
 Y.~Aharanov and L.~Vaidman,
  in \emph{Bohmian Mechanics and Quantum Theory:  An Appraisal},
  J.T.~Cushing, A.~Fine and S.~Goldstein, eds.\ (Kluwer, Dordrecht, 1996).
\bibitem{Scully}
  M.O.~Scully,
  \emph{Physica Scripta T} \textbf{76}, 41 (1998).
\bibitem{Griffiths}
  R.B.~Griffiths,
  Los Alamos e-print, xxx.lanl.gov, quant-ph/9902059.
\bibitem{Fine}
  A.~Fine,
  in \emph{Bohmian Mechanics and Quantum Theory:  An Appraisal},
  J.T.~Cushing, A.~Fine and S.~Goldstein, eds.\ (Kluwer, Dordrecht, 1996).
\bibitem{Bell}
   J.S.~Bell,
    \emph{Speakable and Unspeakable in Quantum Mechanics}
    (Cambridge University Press, Cambridge, 1987).
\bibitem{Page}
  D.N.~Page,
  in \emph{Bohmian Mechanics and Quantum Theory:  An Appraisal},
  J.T.~Cushing, A.~Fine and S.~Goldstein, eds.\ (Kluwer, Dordrecht, 1996).
\bibitem{Squires}
 E.J.~Squires,
   personal communication.
\bibitem{Schrod}
 E.~Schr\"{o}dinger,
 \emph{Naturwissenschaften} \textbf{23}, 807; 823; 844 (1935); reprinted
in 
 J.A.~Wheeler and W.H.~Zurek (eds.), 
 \emph{Quantum Theory and Measurement} (Princeton University Press,
 Princeton N.J., 1983).
\bibitem{GellMann}
  M.~Gell-Mann and J.B.~Hartle,
  \emph{Phys.\ Rev.\ D} \textbf{47}, 3345 (1993).
\bibitem{Omnes}
  R.~Omnes, 
  \emph{The Interpretation of Quantum Mechanics} (Princeton University Press,
  Princeton, 1994).
\bibitem{Joos}
  D.~Giulini, E.~Joos, C.~Kiefer, J.~Kupsch, I.-O.~Stamatescu and H.D.~Zeh,
  \emph{Decoherence and the Appearance of a Classical World 
  in Quantum Theory} 
  (Springer, Berlin, 1996).
\bibitem{Zurek1}
 W.H.~Zurek,
  \emph{Phil.\ Trans.\ Roy.\ Soc.\ Lond.\ A} \textbf{356}, 1793 (1998).
\bibitem{Zeh}
 H.D.~Zeh,
   Los Alamos e-print, xxx.lanl.gov, quant-ph/9812059.
\bibitem{Epstein}
  S.T.~Epstein,
   \emph{Phys.\ Rev.\ }\textbf{89}, 319 (1952); \textbf{91}, 985 (1953).
\bibitem{Bohm3}
  D.~Bohm and J-P.~Vigier,
    \emph{Phys.\ Rev.\ }\textbf{96}, 208 (1954).
\bibitem{Bohm4}
  D.~Bohm and B.J.~Hiley,
    \emph{Phys.\ Rep.\ }\textbf{172}, 93 (1989).
\bibitem{Nel1}
  E.~Nelson,
    \emph{Phys.\ Rev.\ }\textbf{150}, 1079 (1966).
\bibitem{Nel2}
  E.~Nelson,
    \emph{Quantum Fluctuations}
      (Princeton University Press, Princeton N.J., 1985).
\bibitem{Goldstein2}
  S.~Goldstein,
     \emph{J.\ Stat.\ Phys.\ }\textbf{47}, 645 (1987).
\bibitem{Vink}
  J.C.~Vink,
    \emph{Phys.\ Rev.\ A} \textbf{48}, 1808 (1993).
\bibitem{Singh1}
  S.M.~Roy and V.~Singh,
  \emph{Mod.\ Phys.\ Lett.\ A} \textbf{10}, 709 (1995).
\bibitem{Singh2}
  V.~Singh,
  \emph{Pramana} \textbf{49}, 5 (1997).
\bibitem{Singh3}
 S.M.~Roy and V.~Singh,
 Los Alamos e-print, quant-ph/9811041.
\bibitem{Polavieja}
 G.~Garc\'{i}a de Polavieja,
  \emph{Phys.~Lett.~A} \textbf{220}, 303 (1996).
\bibitem{Floyd1}
 E.R.~Floyd, 
 \emph{Found.\ Phys.\ Lett.\ }\textbf{9}, 489 (1996).
\bibitem{Floyd2}
 E.R.~Floyd, 
  \emph{Int.\ J.\ Mod.\ Phys.\ A} \textbf{14}, 1111 (1999).
\bibitem{Faraggi}
 A.E.~Faraggi and M.~Matone,
 Los Alamos e-print, hep-th/9809127.
\bibitem{Bub}
  J.~Bub,
    \emph{Interpreting the Quantum World}
    (Cambridge University Press, Cambridge, 1997).
\bibitem{Suth}
  R.I.~Sutherland,
    \emph{Found.\ Phys.\ }\textbf{27}, 845 (1997).
\bibitem{Ghir}
  E.~Deotto and G.C.~Ghirardi,
     \emph{Found.\ Phys.\ }\textbf{28}, 1 (1998).
\bibitem{Bac2}
  G.~Bacciagaluppi,
  Los Alamos e-print, xxx.lanl.gov, quant-ph/9811040.
\bibitem{Potvin}
 G.~Potvin,
 Los Alamos e-print, xxx.lanl.gov, quant-ph/9908019.
\bibitem{Hus}
  K.~Husimi,
  \emph{Proc.\ Phys.\ Math.\ Soc.\ Jpn.}\ \textbf{22}, 264 (1940).
\bibitem{Hil}
  M.~Hillery, R.F.~O'Connell, M.O.~Scully and E.P.~Wigner,
  \emph{Phys.\ Rep.\ }\textbf{106}, 121 (1984).
\bibitem{Lee}
  H.W.~Lee,
  \emph{Phys.\ Rep.\ }\textbf{259}, 147 (1995).
\bibitem{Arthurs}
  E.~Arthurs and S.C.~Kelly,
  \emph{Bell Syst.\ Tech.\ J.\ }\textbf{44}, 725 (1965).
\bibitem{Ali}
  S.T.~Ali and E.~Prugove\v{c}ki,
    \emph{J.\ Math.\ Phys.\ }\textbf{18}, 219 (1977).
\bibitem{Busch}
  P.~Busch, M.~Grabowski and P.J.~Lahti,
   \emph{Operational Quantum Physics} (Springer, Berlin, 1995).
\bibitem{Leon}
  U.~Leonhardt,
   \emph{Measuring the Quantum State of Light} 
   (Cambridge University Press, Cambridge, 1997).
\bibitem{Self4}
  D.M.~Appleby,
    \emph{Int.\ J.\ Theor.\ Phys.\ }\textbf{37},  2557 (1998).
\bibitem{Self3}
 D.M.~Appleby,
     \emph{Int.\ J.\ Theor.\ Phys.}, \textbf{38}, 807  (1999).
\bibitem{Torres}
  K.B.~M{\o}ller, T.G.~J{\o}rgensen and G.~Torres-Vega,
    \emph{J.~Chem.~Phys.}\ \textbf{106}, 7228 (1997).
\bibitem{Self5}
 D.M.~Appleby,
   \emph{J.\ Mod.\ Opt.}\ \textbf{46}, 825 (1999).
\bibitem{Martin}
 A.~Martin and S.M.~Roy,
  \emph{Phys.\ Lett.\ B}\textbf{350}, 66 (1995).
\end{thebibliography}
\end{document}